\documentclass[11pt,twoside]{article}
\usepackage{graphicx}
\usepackage{amsmath}
\usepackage{amssymb}
\usepackage{mathrsfs}

\begin{document}

\thispagestyle{plain}

\newcommand{\pst}{\hspace*{1.5em}}
\newcommand{\be}{\begin{equation}}
\newcommand{\ee}{\end{equation}}
\newcommand{\ds}{\displaystyle}
\newcommand{\bdm}{\begin{displaymath}}
\newcommand{\edm}{\end{displaymath}}
\newcommand{\bea}{\begin{eqnarray}}
\newcommand{\eea}{\end{eqnarray}}
\newcommand{\bml}{\begin{multline}}
\newcommand{\emltl}{\end{multline}}

\begin{center} {\Large \bf
\begin{tabular}{c}
Pauli equation for joint tomographic probability \\[-1mm]
distribution of spin 1/2 particle
\end{tabular}
 } \end{center}
\smallskip
\begin{center} {\bf Ya. A. Korennoy, V. I. Man'ko}\end{center}
\smallskip
\begin{center}
{\it P.N.    Lebedev Physics Institute,                          \\
       Leninskii prospect 53, 119991, Moscow, Russia }
\end{center}
\begin{abstract}\noindent
The positive vector optical tomogram fully describing the quantum state 
of spin 1/2 particle without any redundancy is introduced.
Reciprocally the vector symplectic tomogram and vector quasidistributions
$\vec W({\mathbf q},{\mathbf p})$, $\vec Q({\mathbf q},{\mathbf p})$,
$\vec P(\vec\alpha)$ are introduced.
The evolution equations for proposed vector optical
and symplectic tomograms and vector quasidistributions for arbitrary Hamiltonian
are obtained. 
The quantum system of charged spin 1/2 particle in arbitrary electro-magnetic field
is considered in proposed representations and evolution equations 
which are analogs of Pauli equation are obtained.
The propagator of evolution equation in the case of
homogeneous and stationary magnetic field in Landau gauge is found and the evolution 
of initial  entangled superposition of lower Landau levels in the vector optical representation 
is considered.
The system of linear quantum oscillator with spin 
in vector optical tomography representation is considered and the evolution of 
initial entangled superposition  of two lower Fock states and spin-up, spin-down states
is studied in this representation.

\end{abstract}


\section{Introduction}
The tomographic approach \cite{BerBer,VogRis} to the quantum state of a system has
allowed one to establish a map between the density operator (or any its representation) and a
set of probability distributions, often called `quantum tomograms'. The latter have all the characteristics
of classical probabilities; they are non-negative, measurable and normalized.

Based on this connection, a classical-like description of quantum dynamics by means of
`symplectic tomography' has been formulated \cite{Mancini96,ManciniFoundPhys97}, providing a bridge between classical and
quantum worlds.  
The tomographic distribution for rotated spin variables
has been constructed in \cite{DodonovManko1997}, and the same approach
has been followed in \cite{Weigert1998}.
Different aspects of classical-like description using
tomographic probabilities were given in 
\cite{IbortPhysScr,MankoMankoFoundPhys2009,LvovRayRevModPhys}.

The main deficiency of the proposed spin tomograms is a redundancy of information.
Attempts to reduce or avoid such a redundancy were made in 
\cite{Fillipov7,Fillipov8,Fillipov9,Fillipov10,Fillipov11,Fillipov12,Fillipov14,Fillipov15}.
The spin tomography was also studied in \cite{FillipovManko2010,MankoMarmo2004}
and in other papers.

The tomographic formulation of quantum evolution equation was suggested in \cite{OlgaJRLR97}
for symplectic tomograms. For optical tomograms it was given in \cite{Korarticle2, Korarticle1}.
The first attempt of foundation of the Pauli equation in tomographic
representation was done in \cite{ManciniOlgaManko2001}.
Using the version of spin tomogram with redundancy of information the authors
obtained a complicated evolution equation.

The aim of our work is the consideration of the special case of spin 1/2 particle quantum state tomography
without redundancy of information, 
constructing the joint vector distribution for space coordinates and spin projections, and
finally deriving the evolution equation for such distribution, which would be an analogue of
the Pauli equation. It would also be a simplification of approach attempted in  
\cite{ManciniOlgaManko2001}.

The paper is organized as follows. 
In Sec. 2 we give  basic formulas of tomographic representation of
quantum mechanics and the evolution equation for optical and symplectic tomogram of nonrelativistic 
spinless quantum system with arbitrary Hamiltonian.
In Sec. 3 we introduce a positive four-component vector probability description of spin 1/2 particle
and give the evolution equation for such a vector-portrait of quantum state with arbitrary Hamiltonian.
In Sec. 4 charged spin 1/2 particle in arbitrary electro-magnetic field
is considered in proposed representations and evolution equations 
which are analogs of Pauli equation are obtained.
In Sec. 5 vector quasidistributions
$\vec W({\mathbf q},{\mathbf p})$, $\vec Q({\mathbf q},{\mathbf p})$,
$\vec P(\vec\alpha)$ are introduced and analogs of Pauli equation for 
$\vec W({\mathbf q},{\mathbf p})$ and $\vec Q({\mathbf q},{\mathbf p})$ 
are obtained.
In Sec. 6 the propagator of evolution equation in the case of
homogeneous and stationary magnetic field in Landau gauge is found and the evolution 
of initial  entangled superposition of lower Landau levels in the vector optical representation 
is considered.
In Sec. 7 the system of linear quantum oscillator with spin 
in vector optical tomography representation is considered and the evolution of 
initial entangled superposition  of two lower Fock states and spin-up, spin-down states
is studied.
The conclusion and prospects are presented in Sec. 8.


\section{Evolution of spinless quantum systems in probability representation}
Let us review the constructions of the optical and symplectic tomograms
for spinless systems. The relationships between the density operator $\hat\rho$ 
and the  optical tomogram $w(\vec X,\vec\theta)$ of the system in the invariant form 
\cite{Korarticle5} are written as follows
\be		\label{equation31_1}
w(\vec X,\vec\theta)=\mbox{Tr}\{\hat \rho\hat U_w(\vec X,\vec\theta)\},~~~
\hat\rho=\int w(\vec X,\vec\theta)\hat D_w(\vec X,\vec\theta) \mbox{d}^nX~\mbox{d}^n\theta,
\ee
where dequantizer $\hat U_w(\vec X,\vec\theta)$ and quantizer $\hat D_w(\vec X,\vec\theta)$ operators equal
respectively
\be		\label{dequantizerOPT}
\hat U_w(\vec X,\vec\theta)=|\vec X,\vec\theta\,\rangle\langle \vec X,\vec\theta\,|=
\hbar^{n/2}\prod_{\sigma=1}^n(m_\sigma\omega_{0\sigma})^{-n/2}
\delta\left(X_\sigma\hat 1-\hat q_\sigma\cos\theta_\sigma-\hat p_\sigma\frac{\sin\theta_\sigma}
{m_\sigma\omega_{0\sigma}}\right),
\ee
\be		\label{quantizerOPT}
\hat D_w(\vec X,\vec\theta)=\left(\frac{\sqrt{\hbar}}{2\pi}\right)^{n}\int\prod_{\sigma=1}^n
\frac{\vert\eta\vert}{\sqrt{m_\sigma\omega_{0\sigma}}}
\exp\left\{i\eta_\sigma\left(X_\sigma-\hat q_\sigma\cos\theta_\sigma
-\hat p_\sigma\frac{\sin\theta_\sigma}
{m_\sigma\omega_{0\sigma}}\right)\right\}
\mbox{d}^n\eta,
\ee
where $|\vec X,\vec\theta\,\rangle$ is an eigenfunction of the operator
$\vec{\hat X}(\vec\theta\,)$ with components
$\hat X_\sigma=\hat q_\sigma\cos\theta_\sigma+\hat p_\sigma\sin\theta_\sigma$
corresponding to the eigenvalue $\vec X$.
Notion of quantizer and dequantizer is related to star product quantization schemes
(see recent review \cite{SIGMA10(2014)086}).

The von-Neumann equation without interaction with the environment 
\be                             \label{vonNeumann}
i\hbar\frac{\partial}{\partial t}\hat\rho=[\hat H,\hat\rho]
\ee
in the optical tomography representation has the form \cite{Korarticle1}
\be                             		\label{r22_2_22222}
\partial_tw({\vec X},{\vec\theta},t)=
\frac{2}{\hbar}\int\mbox{Im}\left[\mbox{Tr}\left\{\hat H\hat D({\vec X}',{\vec\theta}\,')
\hat U({\vec X},{\vec\theta}\,)\right\}\right]w({\vec X}',{\vec\theta}\,',t){\mbox d}^n X'{\mbox d}^n\theta'\,,
\ee
and for a large class of Hamiltonians $\hat H(\hat{\bf p},\hat{\bf q},t)$, when the Hamiltonian
is an analytic function of position $\hat{\bf q}$ and momentum $\hat{\bf p}$ components,
it can be written as follows
\be                             \label{dynEqOpt}
\partial_tw({\vec X},{\vec\theta},t)=
\tilde{\cal M}_w(\vec X,\vec\theta,t)\,w({\vec X},{\vec\theta},t).
\ee
where the operator $\hat{\cal M}(\vec X,\vec\theta,t)$ is obtained form the Hamiltonian
\bdm
\hat{\cal M}_w(\vec X,\vec\theta,t)=
\frac{2}{\hbar}\,\mbox{Im}\,
\hat H\left([\hat{\bf p}]_w(\vec X,\vec\theta\,),[\hat{\bf q}]_w(\vec X,\vec\theta\,),t\right),
\edm
which is an operator depending on two operators of position $\tilde{\bf q}$ and momentum $\tilde{\bf p}$
in the tomographic representation
\be			\label{qFromXTheta}
[\hat q_\sigma]_w(\vec X,\vec\theta)=\sin\theta_\sigma\frac{\partial}{\partial\theta_\sigma}
\left[\frac{\partial}{\partial X_\sigma}\right]^{-1}
+X_\sigma\cos\theta_\sigma+i\frac{\hbar\sin\theta_\sigma}
{2m_\sigma\omega_{\sigma}}
\frac{\partial}{\partial X_\sigma},
\ee
\be			\label{pFromXTheta}
[\hat p_\sigma]_w(\vec X,\vec\theta)=m\omega_{0\sigma}\left(-\cos\theta_\sigma\left[\frac{\partial}{\partial X_\sigma}\right]^{-1}
\frac{\partial}{\partial\theta_\sigma}+X_\sigma\sin\theta_\sigma\right)-\frac{i\hbar}{2}
\cos\theta_\sigma\frac{\partial}{\partial X_\sigma}.
\ee

Similarly, for symplectic tomogram  $M(X,\mu,\nu,t)$
one can be written 
\be		\label{equation31_2}
M(\vec X,\vec\mu,\vec\nu,t)=\mbox{Tr}\{\hat \rho\hat U_M(\vec X,\vec\mu,\vec\nu)\},~~~
\hat\rho=\int M(\vec X,\vec\mu,\vec\nu,t)\hat D_M(\vec X,\vec\mu,\vec\nu) \mbox{d}^nX~\mbox{d}^n\mu~\mbox{d}^n\nu,
\ee
where dequantizer $\hat U_M(\vec X,\vec\mu,\vec\nu)$ and quantizer $\hat D_M(\vec X,\vec\mu,\vec\nu)$
operators are respectively equal
\be		\label{dequantizerSYMP}
\hat U_M(\vec X,\vec\mu,\vec\nu)=|\vec X,\vec\mu,\vec\nu\,\rangle\langle \vec X,\vec\mu,\vec\nu\,|=
\hbar^{n/2}\prod_{\sigma=1}^n(m_\sigma\omega_{0\sigma})^{-n/2}
\delta(X_\sigma\hat 1-\hat q_\sigma\mu_\sigma-\hat p_\sigma\nu_\sigma),
\ee
\be		\label{quantizerSYMP}
\hat D_M(\vec X,\vec\mu,\vec\nu)=\frac{1}{(2\pi\sqrt\hbar)^n}
\prod_{\sigma=1}^n(m_\sigma\omega_{0\sigma})^{3n/2}
\exp\left\{i\sqrt{\frac{m_\sigma\omega_{0\sigma}}{\hbar}}
\left(X_\sigma-\hat q_\sigma\mu_\sigma-\hat p_\sigma\nu_\sigma\right)\right\},
\ee
where $|\vec X,\vec\mu,\vec\nu\,\rangle$ is an eigenfunction of the operator
$\vec{\hat X}(\vec\mu,\vec\nu\,)$ with components
$\hat X_\sigma=\mu_\sigma\hat q_\sigma+\nu_\sigma\hat p_\sigma$
corresponding to the eigenvalue $\vec X$.

For the same Hamiltonians the evolution equation for the symplectic tomogram \cite{Korarticle1}
\be                             \label{dynEqSymp}
\partial_tM({\vec X},{\vec\mu},\vec\nu,t)=
\hat{\cal M}_M({\vec X},{\vec\mu},\vec\nu,t)M({\vec X},{\vec\mu},\vec\nu,t),
\ee
with notation 
\begin{displaymath}
\hat{\cal M}_M({\vec X},{\vec\mu},\vec\nu,t)=\frac{2}{\hbar}\,
\mathrm{Im}\,\hat H\left([\hat{\bf p}]_M({\vec X},{\vec\mu},\vec\nu),[\hat{\bf q}]_M({\vec X}{\vec\mu},\vec\nu),t\right),
\end{displaymath}
where $[\hat{\bf q}]_M$ and $[\hat{\bf p}_M]$ are operators of positions
and momentums in the symplectic representation
\be			\label{defqpSymp}
[\hat p_\sigma]_M=\left(-\left[\frac{\partial}{\partial X_\sigma}\right]^{-1}
\frac{\partial}{\partial\nu_\sigma}-i\frac{\mu_\sigma\hbar}{2}\frac{\partial}{\partial X_\sigma}
\right),~~~~
[\hat q_\sigma]_M=\left(-\left[\frac{\partial}{\partial X_\sigma}\right]^{-1}
\frac{\partial}{\partial\mu_\sigma}+i\frac{\nu_\sigma\hbar}{2}\frac{\partial}{\partial X_\sigma}
\right).
\ee

\section{Probability description of spin 1/2 particle}

As known, that pure  states of quantum spin 1/2 particle
are described by two-component spinor wave functions
$(\psi_1,~\psi_2)$, and mixed states  
can be described by density matrixes $\hat\rho_{ij}$, where $i,\,j=1,\,2$.
In the case of pure state
\be			\label{density}
\hat\rho=\left(\begin{array}{cc}
\psi^*_1\psi_1 & \psi^*_1\psi_2 \\
\psi^*_2\psi_1 & \psi^*_2\psi_2
\end{array}\right).
\ee
The density matrix satisfy Pauli equation (\ref{vonNeumann}) in the von-Neumann form
with the $2\times2$ matrix operator Hamiltonian.
The density matrix is a normalized nonnegative hermitian matrix and thus
it  is actually defined by four real scalar components.
To construct our representation we should pick up four positive components
(spatial distributions), which completely define the density matrix.

Let us choose such components as follows:
\bea
w_1(\vec X,\vec\theta,t)&=&\mbox{Tr}\left\{\hat\rho(t) \hat U_w(\vec X,\vec\theta)
\otimes
|s_1=1/2\rangle\langle s_1=1/2|
\right\}, \nonumber \\[3mm]
w_2(\vec X,\vec\theta,t)&=&\mbox{Tr}\left\{\hat\rho(t) \hat U_w(\vec X,\vec\theta)
\otimes
|s_2=1/2\rangle\langle s_2=1/2|
\right\}, \nonumber \\[3mm]
w_3(\vec X,\vec\theta,t)&=&\mbox{Tr}\left\{\hat\rho(t) \hat U_w(\vec X,\vec\theta)
\otimes
|s_3=1/2\rangle\langle s_3=1/2|
\right\}, \nonumber \\[3mm]
w_4(\vec X,\vec\theta,t)&=&\mbox{Tr}\left\{\hat\rho(t) \hat U_w(\vec X,\vec\theta)
\otimes
|s_3=-1/2\rangle\langle s_3=-1/2|
\right\},
\label{deffN}
\eea
where $\hat U_w(\vec X,\vec\theta)$ is a spinless dequantizer operator (\ref{dequantizerOPT})
and $|s_j=\pm1/2\rangle$ is an  eigenfunction of the projection of spin operator
to the direction $q_j$ corresponding to the eigenvalue $\pm1/2$.
In more compact form formula (\ref{deffN}) can be written as follows
\be			\label{deffn2}
\vec w(\vec X,\vec\theta,t)=\mbox{Tr}\left\{
\hat\rho(t)\vec{\hat U}_w(\vec X,\vec\theta)
\right\},
\ee
where $\vec w(\vec X,\vec\theta,t)$ is a four component vector of probability distributions
and dequantizer operator $\vec{\hat U}_w(\vec X,\vec\theta)$ has the form
\be				\label{dequantoper}
\vec{\hat U}_w(\vec X,\vec\theta)=\hat U_w(\vec X,\vec\theta)
\otimes\vec{\hat{\mathcal U}},
\ee
where $\vec{\hat{\mathcal U}}$ is a four-component vector of $2\times2$ matrices
\be				\label{spinDequv}
\vec{\hat{\mathcal U}}=
\left\{\hat{\mathcal U}_{j(kl)}\right\}=
\left( 
\frac{1}{2}\left[1~~~1 \atop 1~~~1\right],\,\,
\frac{1}{2}\left[1~-i \atop i~~~1\right],\,\,
\left[1~~0 \atop 0~~0\right],\,\,
\left[0~~0 \atop 0~~1\right]
\right).
\ee
Here the first index $j=1,2,3,4$ is the number of the component of the four-component vector,
and $(kl)$ are the indexes of $2\times2$ matrices.
The inverse transform of $\vec w \to \hat\rho$
can be written in terms of quantizer operator $[\vec{\hat D}_{jk}]_w(\vec X,\vec\theta)$
as follows
\be			\label{rhofromF}
\hat\rho_{jk}(t)=\int [\vec{\hat D}_{jk}]_w(\vec X,\vec\theta)\,
\vec w(\vec X,\vec\theta,t)\,\mathrm{d}^3X\mathrm{d}^3\theta\,,~~~~j,k=1,\,2.
\ee
Quantizer operator $[\vec{\hat D}_{jk}]_w(\vec X,\vec\theta)$ in these notations
is defined as a direct product
\be				\label{quantiz}
[\vec{\hat D}_{jk}]_w(\vec X,\vec\theta)=
\hat D_w(\vec X,\vec\theta)\otimes\vec{\hat{\mathcal D}}_{jk}\,\,,
\ee
where $\hat D_w(\vec X,\vec\theta)$ is a spinless quantizer operator (\ref{quantizerOPT})
and $\vec{\hat{\mathcal D}}$ is a $2\times2$ matrix of four-component vectors
\be				\label{spinDopt}
\vec{\hat{\mathcal D}}=
\left\{\hat{\mathcal D}_{(jk)l}\right\}=
\left[
\begin{array}{cc}
\left(0,\,0,\,1,\,0\right) & \left(1,\,-i,\,\frac{-1+i}{2},\,\frac{-1+i}{2}\right) \\
\left(1,\,i,\,-\frac{1+i}{2},\,-\frac{1+i}{2}\right) & \left(0,\,0,\,0,\,1\right)
\end{array}
\right],
\ee
where $(jk)$ are the indexes of $2\times2$ matrix and $l=1,2,3,4$ is the index of the component
of four-component vector.
The functions $w_1(\vec X,\vec\theta)$, $w_2(\vec X,\vec\theta)$,
$w_3(\vec X,\vec\theta)$ are the probability distributions of the operator
$\hat{\vec X}(\vec\theta)$ at time $t$ under the conditions that the particle has the value of spin projection
equal $1/2$ along $q_1$, $q_2$, or $q_3$
directions respectively, and the function $w_4(\vec X,\vec\theta)$ is the 
probability distribution of this operator under the condition that it has the spin projection
$-1/2$ along $q_3$ direction.
Obviously that the two components of the vector $\vec w(\vec X,\vec\theta,t)$
are normalized by the condition
\be			\label{normalization}
\int w_3(\vec X,\vec\theta,t)\mbox{d}^3X+
\int w_4(\vec X,\vec\theta,t)\mbox{d}^3X=1.
\ee
The other two components must be integrable over $X^3$ and
must satisfy the inequalities
\bdm
0\leq w_j(\vec X,\vec\theta,t) \leq 1,~~~~
0\leq \int w_j(\vec X,\vec\theta,t)\mbox{d}^3X \leq 1,~~~~j=1,~2.
\edm
We see, that the four-component vector  $\vec w(\vec X,\vec\theta)$
completely define the density matrix $\hat\rho$ and consequently
it contains all accessible information about the quantum state.

Similarly, for symplectic vector tomography we can write 
\be		\label{deftomogrSymp}
\vec{M}(\vec X,\vec\mu,\vec\nu,t)=\mbox{Tr}\{\hat \rho\vec{\hat U}_M(\vec X,\vec\mu,\vec\nu)\},~~~
\hat\rho_{jk}=\int [\vec{\hat D}_{jk}]_M(\vec X,\vec\mu,\vec\nu)\,
\vec M(\vec X,\vec\mu,\vec\nu,t) \mathrm{d}^3X~\mathrm{d}^3\mu~\mathrm{d}^3\nu,
\ee
where dequantizer $\vec{\hat U}_M(\vec X,\vec\mu,\vec\nu)$ and quantizer
$[\vec{\hat D}_{jk}]_M(\vec X,\vec\mu,\vec\nu)$
are defined by the similar formulas (\ref{dequantoper}) and (\ref{quantiz})
as in the case of optical tomography but the spinless optical dequantizer and quantizer
must be replaced with the corresponding symplectic operators (\ref{dequantizerSYMP})
and (\ref{quantizerSYMP})
\be				\label{DequantQuantSym}
\vec{\hat U}_M(\vec X,\vec\theta)=\hat U_M(\vec X,\vec\theta)
\otimes\vec{\hat{\mathcal U}}\,,
~~~~~~
[\vec{\hat D}_{jk}]_M(\vec X,\vec\theta)=
\hat D_M(\vec X,\vec\theta)\otimes\vec{\hat{\mathcal D}}_{jk}\,\,.
\ee

Generalizing equation (\ref{r22_2_22222}) to the case of spin particles
we can write the evolution equation for the vector tomogram
in spin optical tomography representation
\be                             		\label{r22222VecOpt}
\partial_t\vec w_j({\vec X},{\vec\theta},t)=
\frac{2}{\hbar}\sum_{k=1}^4\int\mathrm{Im}\left[\mathrm{Tr}\left\{\sum_{l,m=1}^2
[\hat U_{j(lm)}]_w({\vec X},{\vec\theta}\,)\hat H\,[\hat D_{(ml)k}]_w({\vec X}',{\vec\theta}\,')
\right\}\right]\vec w_k({\vec X}',{\vec\theta}\,',t){\mbox d}^n X'{\mbox d}^n\theta'\,,
\ee
or in spin symplectic tomography representation
\bea                             		
&&\!\!\!\!\!\!\!\!\!\!\!\!\partial_t\vec w_j(\vec X,\vec\mu,\vec\nu,t)= \nonumber \\[3mm]
&&\!\!\!\!\!\!\!\!\!\!\!\!\frac{2}{\hbar}\sum_{k=1}^4\int\mathrm{Im}\left[\mathrm{Tr}\left\{\sum_{l,m=1}^2
[\hat U_{j(lm)}]_M(\vec X,\vec\mu,\vec\nu\,)\hat H\,[\hat D_{(ml)k}]_M(\vec X',\vec\mu\,',\vec\nu\,')
\right\}\right]\vec w_k(\vec X',\vec\mu\,',\vec\nu\,',t)\mathrm{d}^n X'\mathrm{d}^n\mu'\mathrm{d}^n\nu'\,.
\nonumber \\[0mm]
\label{r22222VecSym}
\eea

\section{Charged spin 1/2 particle in electro-magnetic field}

Let's consider the quantum system of a charged spin 1/2 particle 
with charge $e$, mass $m$ in electro-magnetic field with potentials
${\bf A}({\bf q},t)$, $\varphi({\bf q},t)$. As well known, the Hamiltonian of this system
has the form
\be			\label{Hamiltonian1} 
\hat H=\frac{1}{2m}\left(\hat{\bf p}-\frac{e}{c}{\bf A}\right)^2
+e\varphi-\frac{\varkappa}{s}\hat{\bf s}\,{\bf H}=\hat H_0-\frac{\varkappa}{s}\hat{\bf s}\,{\bf H},
\ee
where $\hat H_0$ is an independent on spin part of Hamiltonian, 
${\bf H}=\mbox{rot}{\bf A}$ is a magnetic field, and $\varkappa$ is a magnetic moment of the particle.

Making the transformation of the Pauli equation (\ref{vonNeumann}) with the 
Hamiltonian (\ref{Hamiltonian1}) with the help of transforms (\ref{deffn2}) and (\ref{rhofromF}) 
we find the evolution equation of the four-component function $\vec w(\vec X,\vec\theta,t)$ dependent on time $t$ and
three component vectors $\vec X$ and $\vec\theta$
\be			\label{maineq}
\partial_t\vec w(\vec X,\vec\theta,t)=
\hat{\mathcal M}_w(\vec X,\vec\theta,t)\,\vec w(\vec X,\vec\theta,t)+
\hat{\mathbf S}_w(\vec X,\vec\theta,t)\,\vec w(\vec X,\vec\theta,t),
\ee
where 
\bdm
\hat{\mathcal M}_w(\vec X,\vec\theta,t)=
\frac{2}{\hbar}\mathrm{Im}\,
\hat H_0\left([\hat{\mathbf q}]_w(\vec X,\vec\theta\,),[\hat{\mathbf p}]_w(\vec X,\vec\theta\,),t\right)
\edm
is an operator depending on two operators of position $[\hat{\mathbf q}]_w$ and momentum $[\hat{\mathbf p}]_w$
defined by (\ref{qFromXTheta}) and (\ref{pFromXTheta}) in the tomographic representation,
and $\hat{\mathbf S}_w(\vec X,\vec\theta,t)$ is a $4\times4$ matrix operator,
responsible for the interaction of spin with the magnetic field.
With omitted arguments and introduced designations
\bdm
[\hat A_j]_w=A_j\left([\hat{\mathbf q}]_w(\vec X,\vec\theta\,),t\right),
~~~~
\tilde{\mathrm H}_j=
[\hat {\mathrm H}_j]_w=\mathrm H_j\left([\hat{\mathbf q}]_w(\vec X,\vec\theta\,),t\right),
\edm
\bdm
\left[\nabla_{\mathbf q}\hat{\mathbf A}\right]_w
=\nabla_{\mathbf q}{\mathbf A}\left(
{\mathbf q}\rightarrow [\hat{\bf q}]_w(\vec X,\vec\theta),t\right)
\edm
the explicit forms of  $\hat{\mathcal M}_w$ and $\hat{\mathbf S}_w$
in general case of timedependent and nonhomogeneous electromagnetic field
are written as
\bea
\hat{\mathcal M}_w(\vec X,\vec\theta,t)&=&
\sum_{n=1}^{3}\omega_{n}\left[\cos^2\theta_n\frac{\partial}{\partial\theta_n}
-\frac{1}{2}\sin2\theta_n\left\{1+X_n\frac{\partial}{\partial X_n}\right\}\right] 
+\frac{2e}{\hbar}\,\mathrm{Im}\,[\hat{\varphi}]_w
\nonumber \\[3mm]
&+&\frac{e^2}{mc^2\hbar}\,\mathrm{Im}[\hat{\bf A}]_w^2
-\frac{2e}{mc\hbar}\,\mathrm{Im}\left[\hat{\mathbf A} \hat{\mathbf p}\right]_w
+\frac{e}{mc}\,\mathrm{Re}\left[{\nabla_{\mathbf q}{\mathbf A}}\right]_w,
\label{explicitM}
\eea
\bdm
[\tilde S_{11}]_w=-2\varkappa\,{\rm Im}\,\tilde{\rm H}_1,~~~~
[\tilde S_{12}]_w=
2\varkappa\left\{-{\rm Im}\,\tilde{\rm H}_2+{\rm Re}\,\tilde{\rm H}_3\right\},
\edm
\bdm
[\tilde S_{13}]_w=
-\varkappa\left\{
2\,{\rm Im}\,\tilde{\rm H}_1
+{\rm Re}\,\tilde{\rm H}_2
-{\rm Im}\,\tilde{\rm H}_2
+{\rm Re}\,\tilde{\rm H}_3
+{\rm Im}\,\tilde{\rm H}_3
\right\},
~~~~
[\tilde S_{14}]_w=
\varkappa\left\{
{\rm Re}\,\tilde{\rm H}_2
+{\rm Im}\,\tilde{\rm H}_2
-{\rm Re}\,\tilde{\rm H}_3
+{\rm Im}\,\tilde{\rm H}_3
\right\},
\edm
\bdm
[\tilde S_{21}]_w=-2\varkappa\left\{{\rm Im}\,\tilde{\rm H}_1+{\rm Re}\,\tilde{\rm H}_3\right\},
~~~~
[\tilde S_{22}]_w=-2\varkappa\,{\rm Im}\,\tilde{\rm H}_2,
\edm
\bdm
[\tilde S_{23}]_w=\varkappa\left\{{\rm Re}\,\tilde{\rm H}_1 + {\rm Im}\,\tilde{\rm H}_1
+{\rm Re}\,\tilde{\rm H}_3 - {\rm Im}\,\tilde{\rm H}_3
\right\},
~~~~
[\tilde S_{24}]_w=\varkappa\left\{-{\rm Re}\,\tilde{\rm H}_1 + {\rm Im}\,\tilde{\rm H}_1
+{\rm Re}\,\tilde{\rm H}_3 + {\rm Im}\,\tilde{\rm H}_3
\right\},
\edm
\bdm
[\tilde S_{31}]_w=-2\varkappa\left\{{\rm Im}\,\tilde{\rm H}_1 - {\rm Re}\,\tilde{\rm H}_2\right\},~~~~
[\tilde S_{32}w]_w=-2\varkappa\left\{{\rm Re}\,\tilde{\rm H}_1 + {\rm Im}\,\tilde{\rm H}_2 \right\},
\edm
\bdm
[\tilde S_{33}]_w=\varkappa\left\{{\rm Re}\,\tilde{\rm H}_1 + {\rm Im}\,\tilde{\rm H}_1
-{\rm Re}\,\tilde{\rm H}_2 + {\rm Im}\,\tilde{\rm H}_2
-2\,{\rm Im}\,\tilde{\rm H}_3
\right\},
~~~~
[\tilde S_{34}]_w=\varkappa\left\{{\rm Re}\,\tilde{\rm H}_1 + {\rm Im}\,\tilde{\rm H}_1
-{\rm Re}\,\tilde{\rm H}_2 + {\rm Im}\,\tilde{\rm H}_2
\right\},
\edm
\bdm
[\tilde S_{41}]_w=-2\varkappa\left\{{\rm Im}\,\tilde{\rm H}_1 + {\rm Re}\,\tilde{\rm H}_2\right\},~~~~
[\tilde S_{42}]_w=2\varkappa\left\{{\rm Re}\,\tilde{\rm H}_1 - {\rm Im}\,\tilde{\rm H}_2 \right\},
\edm
\be			\label{Smatrix}
[\tilde S_{43}]_w=\varkappa\left\{-{\rm Re}\,\tilde{\rm H}_1 + {\rm Im}\,\tilde{\rm H}_1
+{\rm Re}\,\tilde{\rm H}_2 + {\rm Im}\,\tilde{\rm H}_2
\right\},
~~~~
[\tilde S_{44}]_w=\varkappa\left\{-{\rm Re}\,\tilde{\rm H}_1 + {\rm Im}\,\tilde{\rm H}_1
+{\rm Re}\,\tilde{\rm H}_2 + {\rm Im}\,\tilde{\rm H}_2
+2\,{\rm Im}\,\tilde{\rm H}_3
\right\}.
\ee

Making the similar procedure with symplectic vector tomography (\ref{deftomogrSymp})
we can find the evolution equation
\be                             \label{dynEqSympSpin}
\partial_t\vec M({\vec X},{\vec\mu},\vec\nu,t)=
\hat{\mathcal M}_M({\vec X},{\vec\mu},\vec\nu,t)\,\vec M({\vec X},{\vec\mu},\vec\nu,t)
+\hat{\mathbf S}_M(\vec X,\vec\mu,\vec\nu,t)\,\vec M(\vec X,\vec\mu,\vec\nu,t),
\ee
where operator $\hat{\mathcal M}_M({\vec X},{\vec\mu},\vec\nu,t)$ corresponds to spinless part $\hat H_0$
of the Hamiltonian (\ref{Hamiltonian1})
\bea
\hat{\mathcal M}_M({\vec X},{\vec\mu},\vec\nu,t)&=&\frac{2}{\hbar}\,
\mathrm{Im}\,\hat H_0\left([\hat{\mathbf p}]_M({\vec X},{\vec\mu},\vec\nu),
[\hat{\mathbf q}]_M({\vec X}{\vec\mu},\vec\nu),t\right)
=\vec\mu\frac{\partial}{\partial\vec\nu}
+\frac{2e}{\hbar}\,\mathrm{Im}\,[\hat{\varphi}]_M
\nonumber \\[3mm]
&+&\frac{e^2}{mc^2\hbar}\,\mathrm{Im}[\hat{\mathbf A}]_M^2
-\frac{2e}{mc\hbar}\,\mathrm{Im}\left[\hat{\mathbf A} \hat{\mathbf p}\right]_M
+\frac{e}{mc}\,\mathrm{Re}\left[\nabla_{\mathbf q}{\mathbf A}\right]_M,
\eea
where
\bdm
[\hat A_j]_M=A_j\left([\hat{\mathbf q}]_M(\vec X,\vec\mu,\vec\nu\,),t\right),~~~~
[\hat\varphi]_M=\varphi\left([\hat{\mathbf q}]_M(\vec X,\vec\mu,\vec\nu\,),t\right),
\edm
\bdm
[\nabla_{\mathbf q}{\mathbf A}]_M=\nabla_{\mathbf q}{\mathbf A}\left(
{\mathbf q}\rightarrow [\hat{\mathbf q}]_M(\vec X,\vec\mu,\vec\nu\,),t\right), 
\edm
and $[\hat{\mathbf q}]_M$, $[\hat{\mathbf p}]_M$ are position
and momentum operators (\ref{defqpSymp}) in the symplectic representation.
The $4\times4$ matrix operator $\hat{\mathbf S}_M(\vec X,\vec\mu.\vec\nu,t)$ is defined by
the similar  formulae (\ref{Smatrix}) where the operators of components
of the magnetic field $\tilde{\rm H}_j$ must be replaced with corresponding
operators in the symplectic tomography representation
$[\hat{\mathrm H}_j]_M={\mathrm H}_j\left([\hat{\mathbf q}]_M(\vec X,\vec\mu,\vec\nu\,),t\right)$.

\section{Evolution equations of charged spin 1/2 particle 
in Wigner and Husimi representations}
Quasiprobability distributions such as Wigner function or Husimi function
are powerful tools of description of quantum systems. 
For spinless particles their definitions from the density matrix can be written as follows
\be				\label{WfromRho}
W({\mathbf q},{\mathbf p},t)={\mathrm{Tr}}\left\{
\hat\rho(t)\,\hat U_W({\mathbf q},{\mathbf p})\right\},
\ee
\be				\label{QfromRho}
Q({\bf q},{\bf p},t)=Q(\vec\alpha)=\langle\vec\alpha|\hat\rho(t)|\vec\alpha\rangle=
{\mathrm{Tr}}\left\{
\hat\rho(t)\,\hat U_Q({\mathbf q},{\mathbf p})\right\},
\ee
with corresponding "dequantizers"\, having the forms
\be				\label{DeqW}
\hat U_W({\mathbf q},{\mathbf p})=
\frac{1}{(2\pi)^N}\int \left|{\bf q}-{\bf u}/2\rangle
\exp(-i{\bf p}{\bf u}/\hbar)\langle{\bf q}+{\bf u}/2\right|
{\rm d}^Nu,
\ee
\be				\label{DeqQ}
\hat U_Q({\mathbf q},{\mathbf p})=|\vec\alpha\rangle\langle\vec\alpha|,
~~~~\vec\alpha=\frac{1}{\sqrt2}\left(\sqrt{\frac{m\omega}{\hbar}}{\bf q}+
\frac{i}{\sqrt{\hbar m\omega}}{\bf p}\right),
\ee
where $|{\mathbf q}\rangle$ is an eigenvalue of the position operator,
$|\vec\alpha\rangle$ is a coherent state.
Inverse maps  $W\to\hat\rho$ and $Q\to\hat\rho$ are expressed with
corresponding "quantizers"\,
\be			\label{RhofromWigner Husimi}
\hat\rho=\int \hat D_W({\mathbf q},{\mathbf p})\,
W({\mathbf q},{\mathbf p})\,\mathrm{d}^3q\,\mathrm{d}^3p
=\int \hat D_Q({\mathbf q},{\mathbf p})\,
Q({\mathbf q},{\mathbf p})\,\mathrm{d}^3q\,\mathrm{d}^3p
\ee
where quantizers  $\hat D_W$ and $\hat D_Q$ are given by
(see \cite{Wigner32,Husimi40,Mizrahi})
\be			\label{Wigquantizer}
\hat D_W({\mathbf q},{\mathbf p})=2^N
\int{\mathrm d}^Nu\,\exp(2i{\mathbf p}{\mathbf u}/\hbar)|{\mathbf q}+{\mathbf u}\rangle
\langle {\mathbf q}-{\mathbf u}|,
\ee
\bea				
\hat D_Q({\mathbf q},{\mathbf p})&=&\left(\frac{m\omega}{\pi\hbar}\right)^{3/2}
\int{\mathrm d}^3x{\mathrm d}^3y\left\{
|{\mathbf x}\rangle\langle {\mathbf y}|
\exp\left(\frac{m\omega}{2\hbar}({\mathbf x}-{\mathbf y})^2\right)
\right. \nonumber \\[3mm]
&\times &\left.
\exp\left[-\frac{m\omega}{\hbar}\left({\mathbf q}-\frac{{\mathbf x}
+{\mathbf y}}{2}\right)^2
-\frac{m\omega}{\hbar} ({\mathbf x} - {\mathbf y})^2
+\frac{i}{\hbar}{\mathbf p}({\mathbf x} - {\mathbf y})
\right]
\right. \nonumber \\[3mm]
&\times &\left.
\prod_{\sigma=1}^{3}\left[
\sum_{n=0}^{\infty}\frac{(-1)^n}{n!2^n}
H_{2n}\left(\sqrt{\frac{m\omega}{\hbar}}q_\sigma-
\frac{m\omega}{2\hbar}(x_\sigma+y_\sigma)^2\right)
\right]
\right\} 
\label{Husimiquantizer}
\eea
Such definitions provide that the quasidistributions in spinless case
are real functions contrary to density matrices, whose nondiagonal 
elements may be complex. If the particle has spin, the density matrix
$\hat\rho_{jk}$ additionally depend on spin indexes, and usually in 
literature many authors make generalizations of definitions
(\ref{WfromRho}), (\ref{QfromRho}) handling the trace operations as a 
partial trace over all of the variables excepting spin indexes.
In such definitions Wigner function $W{jk}({\bf q},{\bf p},t)$ and
Husimi function $Q{jk}({\bf q},{\bf p},t)$ become 
$(2s+1)\times(2s+1)$ matrices dependent on position and momentum.
But their nondiagonal elements over the spin indexes are not surely real.
So, the main advantage of such quasidistributions with respect to density matrix 
disappears.

To decide this problem let us expand our approach to the quasidistributions
and define four-component Wigner $\vec W({\bf q},{\bf p},t)$ and 
Husimi $\vec Q({\bf q},{\bf p},t)$ vector-functions as follows
\be				\label{vecWfromRho}
\vec W({\mathbf q},{\mathbf p},t)={\mathrm{Tr}}\left\{
\hat\rho(t)\,\vec{\hat U}_W({\mathbf q},{\mathbf p})\right\},
\ee
\be				\label{vecQfromRho}
\vec Q({\bf q},{\bf p},t)={\mathrm{Tr}}\left\{
\hat\rho(t)\,\vec{\hat U}_Q({\mathbf q},{\mathbf p})\right\},
\ee
where dequatizers $\vec{\hat U}_W$ and $\vec{\hat U}_Q$ are defined by the same
formulae as (\ref{dequantoper}) with replacement of $\vec{\hat U}(\vec X,\vec\theta)$
by $\hat U_W({\mathbf q},{\mathbf p})$ or $\hat U_Q({\mathbf q},{\mathbf p})$
\be				\label{DequantWignerHusimi}
\vec{\hat U}_W(\vec X,\vec\theta)=\hat U_W(\vec X,\vec\theta)
\otimes\vec{\hat{\mathcal U}}\,,
~~~~~~
\vec{\hat U}_Q(\vec X,\vec\theta)=\hat U_Q(\vec X,\vec\theta)
\otimes\vec{\hat{\mathcal U}}.
\ee
Such definitions guarantee that all of the components of 
$\vec W({\mathbf q},{\mathbf p},t)$ and $\vec Q({\mathbf q},{\mathbf p},t)$ are real,
more over, all of the components of $\vec Q({\mathbf q},{\mathbf p},t)$ are nonnegative.
Here $W_j({\mathbf q},{\mathbf p},t)$ and $Q_j({\mathbf q},{\mathbf p},t)$ are components of 
Wigner and Husimi vector quasiprobability corresponding definite spin projection along
$q_1$,\, $q_2$, or $q_3$ direction.

Inverse mappings  of (\ref{vecWfromRho}) and (\ref{vecQfromRho}) are obvious
\be
\hat\rho_{jk}(t)=\int [\vec{\hat D}_{jk}]_W({\mathbf q},{\mathbf p})\,
\vec W({\mathbf q},{\mathbf p},t) \mathrm{d}^3q~\mathrm{d}^3p
=\int [\vec{\hat D}_{jk}]_Q({\mathbf q},{\mathbf p})\,
\vec Q({\mathbf q},{\mathbf p},t) \mathrm{d}^3q~\mathrm{d}^3p\,,
\ee
where quantizers $[\vec{\hat D}_{jk}]_W$ and $[\vec{\hat D}_{jk}]_Q$
are defined by the similar formulas as in the case of optical tomography (\ref{dequantoper}), (\ref{quantiz})
but the spinless  optical quantizers
must be replaced with the corresponding operators (\ref{Wigquantizer})
and (\ref{Husimiquantizer})
\be				\label{DequantWigHisSym}
[\vec{\hat D}_{jk}]_W({\mathbf q},{\mathbf p})=
\hat D_W({\mathbf q},{\mathbf p})\otimes\vec{\hat{\mathcal D}}_{jk}\,\,,
~~~~~~
[\vec{\hat D}_{jk}]_Q({\mathbf q},{\mathbf p})=
\hat D_Q({\mathbf q},{\mathbf p})\otimes\vec{\hat{\mathcal D}}_{jk}\,\,.
\ee
Let us give the expression for Wigner function in terms of the Husimi
function
\be				\label{WigInHusimi}
\vec W({\mathbf q},{\mathbf p})=\exp\left(
-\frac{\hbar}{4m\omega}\triangle_{\mathbf q}
-\frac{1}{4m\omega\hbar}\triangle_{\mathbf p}
\right) 
\vec Q({\mathbf q},{\mathbf p}),
\ee
where $\triangle_{\mathbf q}$ and $\triangle_{\mathbf p}$ are Laplace operators
in 3D spaces $\{q_j\}$ and $\{p_j\}$. This formula is a trivial generalization of
the corresponding formula \cite{DavidovichLalovich} for spinless $W$ and $Q$. 

Likewise we can introduce the vector Glauber-Sudarshan P-function \cite{Glauber,Sudarshan}
$\vec P(\vec\alpha,t)$
\bdm
\vec P(\vec\alpha,t)=\mathrm{Tr}\left\{
\hat\rho(t)\vec{\hat U}_P(\vec\alpha)\right\},~~~~
\hat\rho(t)=\int [\vec{\hat D}]_P(\vec\alpha)\,\vec P(\alpha,t)
\mathrm{d}^{2n}\alpha,
\edm
where
\bdm
\vec{\hat U}_P(\vec\alpha)=
\left(\frac{e^{|\vec\alpha|^2}}{\pi^n}
\int|\vec\beta\rangle\langle\beta|e^{|\vec\beta|^2-\vec\beta^*\vec\alpha
+\vec\beta\vec\alpha^*}\mathrm{d}^{2n}\beta\right)
\otimes\vec{\hat{\mathcal U}},
~~~~
[\vec{\hat D}]_P(\vec\alpha)=|\alpha\rangle\langle\alpha|
\otimes\vec{\hat{\mathcal D}}.
\edm

In previous sections we have found the evolution equation of 
charged spin 1/2 particle in electro-magnetic field
in optical and symplectic tomographic representations.
Making similar calculation we can obtain such evolution equation 
for our  vector Wigner function, which will be  a generalization of the
Moyal equation \cite{Moyal1949}
\bea			
\frac{\partial }{\partial t}\vec W({\mathbf q},{\mathbf p},t)&=&
\left[-\frac{\mathbf p}{m}
\frac{\partial}{\partial{\mathbf q}}
+\frac{2e}{\hbar}\,\mbox{Im}\,{\varphi}
\left(
{\bf q}+\frac{i\hbar}{2} \frac{\partial}{\partial {\bf p}},t
\right)
+\frac{e^2}{mc^2\hbar}\,\mathrm{Im}{\bf A}^2
\left(
{\bf q}+\frac{i\hbar}{2} \frac{\partial}{\partial {\bf p}},t
\right)\right.
 \nonumber \\[3mm]
&+&\left.
-\frac{2e}{mc\hbar}\,{\mathrm{Im}}\left\{{\mathbf A}
\left(
{\mathbf q}+\frac{i\hbar}{2} \frac{\partial}{\partial {\mathbf p}},t
\right)
\left(
{\mathbf p}-\frac{i\hbar}{2} \frac{\partial}{\partial {\mathbf q}}
\right)
\right\}\right.  \nonumber \\[3mm]
&+&\left.\frac{e}{mc}\,\mathrm{Re}{\nabla_{\bf q}{\bf A}}
\left(
\mathbf q \to {\mathbf q}+\frac{i\hbar}{2} \frac{\partial}{\partial {\mathbf p}},t
\right)
+\hat{\mathbf S}_W({\bf q},{\bf p},t)
 \right] \vec W({\mathbf q},{\mathbf p},t),
\label{Moyal2}
\eea
where $4\times4$ matrix operator $\hat{\mathbf S}_W({\bf q},{\bf p},t)$
is defined by the same formulae 
(\ref{Smatrix}) where the operators of components
of the magnetic field $\tilde{\rm H}_j$ must be replaced with corresponding
operators in the Wigner representation
${\mathrm H}_j\left( {\mathbf q}+\frac{i\hbar}{2} \frac{\partial}{\partial {\mathbf p}},t \right)$.

The corresponding equation for Husimi function is obtained from (\ref{Moyal2}) 
with the help of expression (\ref{WigInHusimi}) (see also \cite{Mizrahi}). For simplicity we choose 
the system of measurements so that $m=\omega=\hbar=1$
\bea			
\frac{\partial }{\partial t}\vec Q({\mathbf q},{\mathbf p},t)&=&
\left[-{\mathbf p}
\frac{\partial}{\partial{\mathbf q}}-\frac{1}{2}\frac{\partial}{\partial{\mathbf q}}
\frac{\partial}{\partial{\mathbf p}}
+\frac{2e}{\hbar}\,\mbox{Im}\,{\varphi}
\left(
{\bf q}+\frac{1}{2}\frac{\partial}{\partial{\mathbf q}}
+\frac{i}{2} \frac{\partial}{\partial {\bf p}},t
\right)\right.  \nonumber \\[3mm]
&+&\frac{e^2}{c^2}\,\mathrm{Im}{\bf A}^2
\left(
{\bf q}
+\frac{1}{2}\frac{\partial}{\partial{\mathbf q}}
+\frac{i}{2} \frac{\partial}{\partial {\bf p}},t
\right)
 \nonumber \\[3mm]
&-&\frac{2e}{c}\,{\mathrm{Im}}\left\{{\mathbf A}
\left(
{\mathbf q}
+\frac{1}{2}\frac{\partial}{\partial{\mathbf q}}
+\frac{i}{2} \frac{\partial}{\partial {\mathbf p}},t
\right)
\left(
{\mathbf p}+\frac{1}{2}\frac{\partial}{\partial{\mathbf p}}
-\frac{i}{2} \frac{\partial}{\partial {\mathbf q}}
\right)
\right\}  \nonumber \\[3mm]
&+&\left.\frac{e}{c}\,\mathrm{Re}{\nabla_{\bf q}{\bf A}}
\left(
\mathbf q \to {\mathbf q}
+\frac{1}{2}\frac{\partial}{\partial{\mathbf q}}
+\frac{i}{2} \frac{\partial}{\partial {\mathbf p}},t
\right)+\hat{\mathbf S}_Q({\bf q},{\bf p},t)
 \right] \vec Q({\mathbf q},{\mathbf p},t),
\label{MoyalHusimi}
\eea
where $4\times4$ matrix operator $\hat{\mathbf S}_Q({\bf q},{\bf p},t)$
is defined by (\ref{Smatrix}) in which components
of the magnetic field $\tilde{\rm H}_j$ are  replaced with 
${\mathrm H}_j\left( {\mathbf q}
+\frac{1}{2}\frac{\partial}{\partial{\mathbf q}}
+\frac{i}{2} \frac{\partial}{\partial {\mathbf p}},t \right)$.

\section{Charged spin 1/2 in homogeneous magnetic field}
Choose the vector and scalar potentials as follows (Landau gauge)
\be			\label{statmagn}
{\bf A}=(-q_2\mathrm H,0,0),~~~~\varphi=0,~~~~{\mathbf H}=(0,\,0,\,\mathrm H),
\ee
suppose $\omega_1=\omega_2=\omega_3=|\omega|$,
where $\omega=\ds\frac{e\mathrm H}{mc}$,
and choose the system of measurements so that 
$|\omega|=m=\hbar=1$. In these units $\omega=+1$ when the charge is positive and $\omega=-1$
when it is negative. 

Also denote the frequency of spin rotation as $\omega_0=\ds\frac{\varkappa\mathrm H}{\hbar}$.
For  electrons $\omega_0=\omega$, but for other particles these two frequencies may be different.

Then the Hamiltonian will have the form
\be			\label{statmagHam}
\hat H=\frac{1}{2}\hat p_1^2+\frac{1}{2}\hat p_2^2+
\frac{1}{2}\hat p_3^2+\frac{1}{2}\hat q_2^2
+\omega\hat p_1\hat q_2  -\frac{\omega_0}{2}\left(1~~~~0\atop 0~-1\right).
\ee
Using the general formulae (\ref{maineq}) -- (\ref{Smatrix})  we find the evolution equation 
\bea
\frac{\partial}{\partial t}\vec w(\vec X,\vec\theta,t)&=&
\left[\sum_{\sigma=1,\,3}\left(\cos^2\theta_\sigma\frac{\partial}{\partial{\theta_\sigma}}
-\frac{1}{2}\sin2\theta_\sigma\left\{1+X_\sigma\frac{\partial}{\partial{X_\sigma}}\right\}\right)
+\frac{\partial}{\partial{\theta_2}} \right.
\nonumber \\[3mm]
&-&\omega\left(\cos\theta_1\left[\frac{\partial}{\partial{X_1}}\right]^{-1}
-X_1\sin\theta_1\right)\sin\theta_2\frac{\partial}{\partial{X_2}}
\nonumber \\[3mm]
&-&\omega\left.\left(\sin\theta_2\left[\frac{\partial}{\partial{X_2}}\right]^{-1}
+X_2\cos\theta_2\right)\cos\theta_1\frac{\partial}{\partial{X_1}}
+\hat{\mathbf S}_w\right]\vec w(\vec X,\vec\theta,t)\,,
\label{evoleqFnstatH}
\eea
where upper signs correspond to positive charge and lower signs
correspond to negative charge, and the matrix $\hat{\mathbf S}_w$
is given by the expression
\be			\label{SmatrixStatH}
\hat{\mathbf S}_w=\omega_0\left[
\begin{array}{cccc}
 0 & 1 & -1/2 & -1/2 \\
-1 & 0 &  1/2 &  1/2 \\
 0 & 0 &    0  &   0   \\
 0 & 0 &    0  &   0   
\end{array}
\right]\,,
\ee
or in symplectic tomography representation
\bea			
\frac{\partial}{\partial t}\vec M(\vec X,\vec\mu,\vec\nu)&=&
\left[
\vec\mu\frac{\partial}{\partial\vec\nu}
-\omega\nu_2\left[\frac{\partial}{\partial X_1}\right]^{-1}
\frac{\partial}{\partial X_2}\frac{\partial}{\partial \nu_1}\right.
 \nonumber \\[3mm]
&+&\left.\omega\mu_1\left[\frac{\partial}{\partial X_2}\right]^{-1}
\frac{\partial}{\partial X_1}\frac{\partial}{\partial \nu_2}
+\hat{\mathbf S}_M \right]
\vec M(\vec X,\vec\mu,\vec\nu)\,.
\label{SymevoleqFnstatH}
\eea
In Wigner representation this equation has the form
\be			\label{PauliWigHomH}
\frac{\partial}{\partial t}\vec W(\mathbf{q},\mathbf {p},t)=\left[
-\mathbf p\frac{\partial}{\partial\mathbf q}
+q_2\frac{\partial}{\partial p_2}
+\omega p_1\frac{\partial}{\partial p_2}
-\omega q_2\frac{\partial}{\partial q_1}
+\hat{\mathbf S}_w
\right]\vec W(\mathbf{q},\mathbf {p},t)\,.
\ee
It is obvious that for homogeneous magnetic field 
$\hat{\mathbf S}_M=\hat{\mathbf S}_w=\hat{\mathbf S}_W=\hat{\mathbf S}_Q$.

If we integrate equation (\ref{evoleqFnstatH}) over $\mathrm d^3X$
and introduce the notation 
$\int\vec w(\vec X,\vec\theta,t)\mathrm d^3X=\vec P(t)$,
then equation (\ref{evoleqFnstatH}) take the form
\be
\partial_t\vec P(t)=\hat{\mathbf S}_w\vec P(t).
\ee
This equation corresponds to the case when we are interesting in only spin dynamic.
After some calculations we get the propagator of this equation and, consequently,
the solution $\vec P(t)=\hat\Pi_s(t)\vec P(0)$ for arbitrary initial condition $\vec P(0)$
\be				\label{PropagatorSpin}
\hat\Pi_s(t)=
\left[
\begin{array}{cccc}
\cos\omega_0 t       & \sin\omega_0 t & (1-\cos\omega_0 t - \sin\omega_0 t)/2 & 
(1-\cos\omega_0 t - \sin\omega_0 t)/2 \\
-\sin\omega_0 t & \cos\omega_0 t     & (1-\cos\omega_0 t + \sin\omega_0 t)/2 & 
(1-\cos\omega_0 t + \sin\omega_0 t)/2 \\
 0 & 0 &    1  &   0   \\
 0 & 0 &    0  &   1   
\end{array}
\right]\,,
\ee
We can see, that the probability of finding the particle in the state with the 
spin projection $\pm1/2$  along $q_3$ direction remains constant during evolution.

The spinless part of the propagator for eq. (\ref{evoleqFnstatH}) or (\ref{SymevoleqFnstatH})
corresponds to the free motion along $q_3$ direction and to the evolution of ordinary
quadratic system with respect to $q_1$ and $q_2$ degrees of freedom.
For free motion the propagator was found in \cite{Korarticle4}
for optical tomography and in \cite{OlgaJRLR97} for symplectic one
\be			\label{PropfreeOpt}
[\Pi_{\mathrm f}]_{w}(X_3,\theta_3,X_3',\theta_3',t)=
\delta(X_3\cos\theta_3'-X_3'\cos\theta_3)
\delta(\cos\theta_3'(t+\tan\theta_3)-\sin\theta_3'),
\ee
\be			\label{PropfreeSym}
[\Pi_{\mathrm f}]_{M}(X_3,\mu_3,\nu_3,X_3',\mu_3',\nu_3')=
\delta(X_3-X_3')\delta(\nu_3'-\nu_3-\mu_3t)\delta(\mu_3-\mu_3').
\ee
Free motion propagator for the Wigner function, obviously, equals
\be			\label{PropFreeWig}
[\Pi_{\mathrm f}]_W(q_3,p_3,q_3',p_3',t)=\delta\left(q_3-q_3'-\frac{p_3'}{m}t\right)
\delta(p_3-p_3').
\ee
For quadratic subsystem the propagator
can be found by the method of motion integrals,
or it can be obtained from the known propagator for the wave function
by means of transformation to the corresponding representation
(see \cite{Korarticle4, Laser41, MankoRosaVitale}).
After some calculations we have
\bea
&&[\Pi_{12}]_w(X_1,X_2,\theta_1,\theta_2,X_1',X_2',\theta_1',\theta_2',t)=
\nonumber \\[3mm]
&&~~~~~~=\delta\left\{
\cos\theta_1\tan\left(\frac{\omega t}{2}\right)[2\cos\theta_1'-\omega\sin\theta_2']
+\sin(\theta_1-\theta_1')
\right\} \nonumber \\[3mm]
&&~~~~~~\times\delta\left\{
\cos\theta_2\tan\left(\frac{\omega t}{2}\right)\left[
2\cos\theta_2'+\omega\frac{\sin(\theta_1-\theta_1')}{\cos\theta_1\sin\theta_1}
\right]
+\sin(\theta_2-\theta_2')
+\omega\frac{\sin\theta_2\sin(\theta_1-\theta_1')}{\cos\theta_1\sin\theta_1}
\right\}\nonumber \\[3mm]
&&~~~~~~\times\delta\left\{X_1\frac{\cos\theta_1'}{\cos\theta_1}
+\omega X_2\frac{\sin(\theta_1-\theta_1')}{\cos\theta_2\cos\theta_1\sin\theta_1}
-X_1'\right\}\,
\delta\left\{X_2\frac{\cos\theta_2'}{\cos\theta_2}-X_2'\right\},
\label{PropHomHOpt}
\eea
or in symplectic representation
\bea
&&[\Pi_{12}]_M(X_1,X_2,\mu_1,\mu_2,\nu_1,\nu_2,X_1',X_2',\mu_1',\mu_2',\nu_1',\nu_2',t)=
\nonumber \\[3mm]
&&~~~~~~=(2\pi)^{-2}\exp\left\{
i\left[-X_1\frac{\mu_1'}{\mu_1}-X_2\frac{\mu_2'}{\mu_2}+X_1'+X_2'
-\frac{\omega X_2}{\mu_2}\left(\frac{\mu_1'}{\mu_1}-\frac{\nu_1'}{\nu_1}\right)
\right]
\right\}\nonumber \\[3mm]
&&~~~~~~\times\delta\left\{\mu_1\tan\left(\frac{\omega t}{2}\right)
(2\mu_1'-\omega\nu_2')-\mu_1\nu_1'+\mu_1'\nu_1
\right\}\nonumber \\[3mm]
&&~~~~~~\times\delta\left\{\mu_2\tan\left(\frac{\omega t}{2}\right)
\left[2\mu_2'+\omega\left(\frac{\mu_1'}{\mu_1}-\frac{\nu_1'}{\nu_1}\right)\right]
+\omega\nu_2\left(\frac{\mu_1'}{\mu_1}-\frac{\nu_1'}{\nu_1}\right)
-\mu_2\nu_2'+\mu_2'\nu_2
\right\}.
\label{PropHomHSymp}
\eea
In Wigner representation this propagator has the form
\bea
&&[\Pi_{12}]_W(q_1,q_2,p_1,p_2,q_1',q_2',p_1',p_2',t)=
\frac{m^2\omega^2}{4\hbar^2\sin^2(\omega t/2)}
\nonumber \\[3mm]
&&~~~~~~
\times\delta\left\{\frac{m\omega}{2\hbar}\cot\left(\frac{\omega t}{2}\right)
(q_1-q_1')-\frac{m\omega}{2\hbar}(q_2-q_2')-p_1'
\right\}\nonumber \\[3mm]
&&~~~~~~\times\delta\left\{\frac{m\omega}{2\hbar}\cot\left(\frac{\omega t}{2}\right)
(q_2-q_2')+\frac{m\omega}{2\hbar}(q_1-q_1')-p_2'
\right\}
\nonumber \\[3mm]
&&~~~~~~\times\delta\left\{\frac{m\omega}{\hbar}
(q_1-q_1')+p_2-p_2'
\right\}
\delta\left\{ p_1-p_1' \right\}.
\label{PropHomHWig}
\eea
The total propagator of eq.(\ref{evoleqFnstatH}) equals to the product
of corresponding propagators (\ref{PropagatorSpin}), (\ref{PropfreeOpt}),
and (\ref{PropHomHOpt})
\be			\label{TotPropOpt}
\hat\Pi_w(\vec X,\vec\theta,\vec X',\vec\theta\,',t)=
\hat\Pi_s\otimes[\Pi_{\mathrm f}]_w\otimes[\Pi_{12}]_w.
\ee
Reciprocally for symplectic or Wigner representation we have
\be			\label{TotPropSymp}
\hat\Pi_M(\vec X,\vec\mu,\vec\nu,\vec X',\vec\mu\,',\vec\nu\,',t)=
\hat\Pi_s\otimes[\Pi_{\mathrm f}]_M\otimes[\Pi_{12}]_M\,,
\ee
\be			\label{TotPropWig}
\hat\Pi_W({\mathbf q},{\mathbf p},{\mathbf q}',{\mathbf p}',t)=
\hat\Pi_s\otimes[\Pi_{\mathrm f}]_W\otimes[\Pi_{12}]_W\,.
\ee
Let us  average the evolution of the system over free motion along $q_3$ direction, 
i.e. integrate the evolution equation and initial condition over $X_3$ and 
consider an initial condition which is the entangled superposition
of lower Landau levels of electron  in Landau gauge (\ref{statmagn}) (the charge is negative 
and in our notations $\omega =\omega_0=-1$).
For this we introduce two pairs of creation and annihilation
operators $\hat a$, $\hat a^\dagger$ and $\hat b$, $\hat b^\dagger$
(see \cite{LandauLifzhitz,MalkinManko})
\be			\label{hata}
\hat a=(\hat p_1-\hat q_2-i\hat p_2) /\sqrt{2},~~~~
\hat a^\dagger=(\hat p_1-\hat q_2+i\hat p_2) /\sqrt{2},
\ee
\be			\label{hatb}
\hat b=(\hat q_1-\hat p_2+i\hat p_1) /\sqrt{2},~~~~
\hat b^\dagger=(\hat q_1-\hat p_2-i\hat p_1) /\sqrt{2}.
\ee
These operators have the following properties
\bdm
[\hat a,\hat a^\dagger]=1,~~~~[\hat b,\hat b^\dagger]=1,
~~~~[\hat a,\hat b]=[\hat a,\hat b^\dagger]=0,
\edm
and Hamiltonian (\ref{statmagHam}) without $q_3$ degree of freedom is expressed
in terms of $\hat a$ and $\hat a^\dagger$ as
\bdm
\hat H=\hat a^\dagger\hat a+\frac{1}{2} +
\frac{1}{2}\left(1~~~~0\atop 0~-1\right).
\edm
The Landau levels correspond to the states with wave functions
\bdm
|nm\rangle=\frac{(\hat a^\dagger)^n(\hat b^\dagger)^m}{\sqrt{n!m!}}|0\,0\rangle,
\edm
where $|0\,0\rangle$ is a vacuum state of the system
\bdm
\hat a|0\,0\rangle=0,~~~~\hat b|0\,0\rangle=0,~~~~
\langle 0\,0|0\,0\rangle=1,
\edm
\bdm
\langle q_1,q_2|0\,0\rangle=\frac{1}{\sqrt{2\pi}}\exp\left(
-\frac{q_1^2}{4}-\frac{q_2^2}{4}+i\frac{q_1q_2}{2}
\right).
\edm
The first exited state $|1\,0\rangle$ has the wave function
\bdm
\langle q_1,q_2|1\,0\rangle=
\langle q_1,q_2|\hat a^\dagger|0\,0\rangle=\frac{iq_1-q_2}{2\sqrt{\pi}}\exp\left(
-\frac{q_1^2}{4}-\frac{q_2^2}{4}+i\frac{q_1q_2}{2}
\right).
\edm
Consider an initial condition which is the entangled superposition
of lower Landau levels
\bdm
|\Psi(0)\rangle=\frac{1}{\sqrt2}(|0\,0\rangle\otimes|-1/2\rangle
+|1\,0\rangle\otimes|1/2\rangle).
\edm
It corresponds to our vector optical tomogram
\bdm
\vec w(X_1,X_2,\theta_1,\theta_2,0)=\frac{1}{4}\left(
\begin{array}{c}
w_{0000}+w_{1010}+2\mathrm{Re}\,w_{0010} \\[-2mm]
w_{0000}+w_{1010}+2\mathrm{Im}\,w_{0010} \\[-2mm]
2w_{1010} \\[-2mm]
2w_{0000}
\end{array}
\right),
\edm
where we introduce the designation
\bdm
w_{nmn'm'}=\langle X_1,X_2,\theta_1,\theta_2|nm\rangle
\langle n'm'|X_1,X_2,\theta_1,\theta_2\rangle.
\edm
It is easy to see that the corresponding solution of evolution equation will be
\bdm
\vec w(X_1,X_2,\theta_1,\theta_2,t)=\frac{1}{4}\left(
\begin{array}{c}
w_{0000}+w_{1010}+2\mathrm{Re}\{w_{0010}\exp(i2t)\} \\[-2mm]
w_{0000}+w_{1010}+2\mathrm{Im}\{w_{0010}\exp(i2t)\} \\[-2mm]
2w_{1010} \\[-2mm]
2w_{0000}
\end{array}
\right).
\edm
Double frequency here is the result of interference of simultaneous
spin rotation and  cyclotron quantum motion in the plane perpendicular
to the magnetic field.

\section{Linear harmonic oscillator with spin}

As another example we consider a system with following Hamiltonian:
\bdm
\hat H=\frac{1}{2}(p^2+q^2)+\left(1~~~~0\atop 0~-1\right).
\edm
It could describe one vibrational degree of a trapped electron plus its spin
\cite{BrownGabrielse}. The measurability of tomograms in this system was
investigated in \cite{MassiniFortunato}.

The evolution equation of the vector tomogram for this system has the simple form
\be				\label{evolsimp}
\frac{\partial}{\partial t}\vec w(X,\theta,t)=
\frac{\partial}{\partial\theta}\vec w(X,\theta,t)+\hat{\mathbf S}_w\vec w(X,\theta,t),
\ee
where $\hat{\mathbf S}_w$ is defined by (\ref{SmatrixStatH}) with $\omega_0=-2$,
and the propagator for this equation be
\bdm
\hat\Pi(X,\theta,t)=\delta(\theta-t-\theta')\otimes\hat\Pi_s.
\edm
If we take an initial entangled state
\bdm
|\Psi(0)\rangle=\frac{1}{\sqrt{2}}(|0\rangle\otimes|-1/2\rangle
+|1\rangle\otimes|1/2\rangle)
\edm
with initial vector optical tomogram
\bdm
\vec w(X,\theta,0)=\frac{1}{4}\left(
\begin{array}{c}
w_{00}+w_{11}+2\mathrm{Re}\,w_{01} \\[-2mm]
w_{00}+w_{11}+2\mathrm{Im}\,w_{01} \\[-2mm]
2w_{11} \\[-2mm]
2w_{00}
\end{array}
\right),
\edm
where
\bea
w_{00}=\langle X,\theta|0\rangle\langle0|X,\theta\rangle&=&
\frac{1}{\sqrt{\pi}}e^{-X^2},\nonumber \\[3mm]
w_{11}=\langle X,\theta|1\rangle\langle1|X,\theta\rangle&=&
\frac{2}{\sqrt{\pi}}X^2e^{-X^2},\nonumber \\[3mm]
w_{01}=\langle X,\theta|1\rangle\langle1|X,\theta\rangle&=&
\frac{\sqrt{2}}{\sqrt{\pi}}Xe^{i\theta}e^{-X^2},\nonumber
\eea
then, the solution of equation (\ref{evolsimp}) be
\bdm
\vec w(X,\theta,t)=\frac{1}{4}\left(
\begin{array}{c}
w_{00}+w_{11}+2\mathrm{Re}\{w_{01}\exp(i3t)\} \\[-2mm]
w_{00}+w_{11}+2\mathrm{Im}\{w_{01}\exp(i3t)\} \\[-2mm]
2w_{11} \\[-2mm]
2w_{00}
\end{array}
\right),
\edm
and we can see again the addition of two frequencies: the harmonic oscillator frequency
and the frequency of spin rotation.

\section{Conclusion}
To resume we point out the main results of our paper.
We suggested to describe the state of charged spin 1/2 particle by a new four-component
positive vector of joint probability distributions, that is the vector optical
tomogram. 
Such approach of construction of positive vector-portrait of quantum 
state eliminates the redundancy, which is the main difficulty
of schemes proposed by another authors.
Reciprocally we introduce the vector symplectic tomogram and vector quasidistributions
$\vec W({\mathbf q},{\mathbf p})$, $\vec Q({\mathbf q},{\mathbf p})$,
$\vec P(\vec\alpha)$.

We obtained the evolution equations for such vector optical
and symplectic tomograms and vector quasidistributions for arbitrary Hamiltonian.
We considered in proposed representations the quantum system of 
charged spin 1/2 particle in arbitrary electro-magnetic field and obtained evolution equations,
which are analogs of Pauli equation in appropriate 
representations.

As an example we found the propagator of evolution equation in the case of
homogeneous and stationary magnetic field in Landau gauge, we considered the evolution 
of initial  entangled superposition of lower Landau levels in the vector optical representation
and illustrated  the addition of the frequency of simultaneous
spin rotation and  the frequency of cyclotron quantum motion in the plane perpendicular
to the magnetic field.

Also as an example we considered the system of linear quantum oscillator with spin 
in vector optical tomography representation and studied the evolution of 
initial entangled superposition  of two lower Fock states and spin-up spin-down states.

A possible disadvantage of the approach proposed is a relatively
complicated evolution equations, but this is the price one ought to pay for the possibility
of describing quantum objects in term of classical probabilities.
In addition, the equations obtained in this paper are much more easier than in the 
previous attempt \cite {ManciniOlgaManko2001} of description of evolution of spin 
particles in terms of probabilities.

The generalization of the results of this paper to the higher spin particles 
will be given in further publications.


\end{document}